\documentclass{article}
\usepackage{amsmath}
\usepackage{graphicx}
\usepackage{float}
\usepackage{natbib}
\usepackage{url} 
\usepackage{graphicx}
\usepackage{amsthm,amsmath,amsfonts,amssymb}
\usepackage{natbib}
\usepackage{amscd,amssymb,verbatim}
\usepackage{mathrsfs}
\usepackage{bbm}
\usepackage{listings}
\usepackage{epsfig}
\usepackage{enumitem}
\usepackage{subcaption}
\usepackage[plain]{algorithm2e}
\usepackage{algorithmic}
\usepackage{url}
\usepackage{booktabs,makecell,ltablex}
\usepackage{multirow}
\usepackage{moreverb}
\usepackage{xcolor}
\usepackage{setspace}
\usepackage{subcaption} 
\usepackage[all,cmtip]{xy}
\usepackage{rotating}

\graphicspath{{Figs/}}
\makeatother

\begin{document}

\def\spacingset#1{\renewcommand{\baselinestretch}%
{#1}\small\normalsize} \spacingset{1}


\title{\bf Bayesian Sensitivity Analysis for Causal Estimation with Time-varying Unmeasured Confounding}
\author{Yushu Zou\thanks{Dalla Lana School of Public Health, University of Toronto, Ontario, Canada and Public Health Ontario, Ontario, Canada.}, \hspace{.2cm}
	Liangyuan Hu\thanks{Rutgers School of Public Health, Rutgers University, New Jersey, United States.}, \hspace{.2cm} Amanda Ricciuto\thanks{Division of Gastroenterology, Hepatology Nutrition, The Hospital for Sick Children, Ontario, Canada.}, \hspace{.2cm} Mark Deneau\thanks{University of Utah and Intermountain
		Healthcare Primary Children's Hospital, Utah, United States.}, \hspace{.2cm} Kuan Liu\thanks{Institute of Health Policy, Management and Evaluation, University of Toronto, Ontario, Canada.} \hspace{.2cm} \\}
 \date{ }
  \maketitle

\bigskip
\begin{abstract}
Causal inference relies on the untestable assumption of no unmeasured confounding. Sensitivity analysis can be used to quantify the impact of unmeasured confounding on causal estimates. Among sensitivity analysis methods proposed in the literature for unmeasured confounding, the latent confounder approach is favoured for its intuitive interpretation via the use of bias parameters to specify the relationship between the observed and unobserved variables and the sensitivity function approach directly characterizes the net causal effect of the unmeasured confounding without explicitly introducing latent variables to the causal models. In this paper, we developed and extended two sensitivity analysis approaches, namely the Bayesian sensitivity analysis with latent confounding variables and the Bayesian sensitivity function approach for the estimation of time-varying treatment effects with longitudinal observational data subjected to time-varying unmeasured confounding. We investigated the performance of these methods in a series of simulation studies and applied them to a multi-center pediatric disease registry data to provide practical guidance on their implementation.
\end{abstract}

\noindent%
\vfill

\newpage
\spacingset{1.5}

\section{Introduction}\label{sec1}

Observational studies provide a feasible, efficient, and cost-effective design for gathering evidence to study treatment and exposure effects.\citep{liu2021pilot} These data present inherent complexities for comparative effectiveness research, such as time-varying treatment-confounding feedback, where the confounders change over time and are influenced by past treatment. Administrative databases are a common source of observational data. Although administrative data are rich in information that spans diagnostic codes, health service billing codes, prescription records, and other information collected at each health services encounter, key variables such as detailed clinical biomarkers and sociodemographic variables (e.g., education level and household income) are not always captured.\citep{blair2021identifying} These variables are often considered unmeasured confounding variables in the causal conceptual framework. For causal estimation, unmeasured confounding violates the strongly ignorable treatment assignment assumption and can lead to biased estimates of treatment effects.\citep{rubin1980randomization} 

Sensitivity analysis can be applied to quantify the impact of unmeasured confounding on the causal estimation, and has been advocated by the Strengthening the Reporting of Observational Studies in Epidemiology (STROBE) guidelines for the purpose of examining the influence of potential unmeasured confounding.\citep{von2007strobe}  Unmeasured confounding is typically captured through one or more non-identifiable numerical parameters, commonly termed bias parameters.\citep{greenland2005multiple, brumback2004sensitivity} One proceeds by specifying and formulating the bias parameters with a plausible range of values and derives the bias-corrected causal estimand using these bias parameters. There are two streams of sensitivity analysis approaches that have been proposed under this framework, namely the latent variable approach and the sensitivity function approach (also known as the confounding function approach).  

The latent variable approach introduces the unmeasured confounding as a single latent/unmeasured variable in the causal framework. Rosenbaum and Rubin first proposed this approach to examine how sensitive the conclusions of a study with a binary outcome are to a binary unmeasured confounder.\citep{Rosenbaum1983AssessingOutcome}  Several sensitivity methods have been developed under the latent variable approach for cross-sectional data, including parametric,\citep{phillips2003quantifying, steenland2004monte, McCandless2008AConfounding, McCandless2017AConfounding, McCandless2019BayesianAnalysis, rose2023monte} non-parametric,\citep{Franks2020FlexibleImplications} and semi-parametric.\citep{Zhang2022AStudies, Dorie2016AConfounding} Many approaches with the latent confounding variables fall under probabilistic sensitivity analysis, which posit probability distributions for the bias parameters and average over the distributions to obtain bias-adjusted estimates, and can be estimated under full Bayesian inference or Monte Carlo estimation. The latent variable approach is favoured when external knowledge or external data is available to inform the distribution of the unmeasured confounding and the strength of association between the unmeasured confounding and the measured variables. 

The sensitivity function approach is framed based on the exchangeability assumption, it uses the sensitivity function to characterize the net causal effect influenced by the unmeasured confounding. The sensitivity function is viewed as the bias parameter and used to derive bias-corrected estimations of the average potential outcome to quantify the influence of unmeasured confounding on the average treatment effect. Existing methods for the sensitivity function approach include parametric,\citep{brumback2004sensitivity, Li2011PropensityConfounding, Shardell2018JointData} and semi-parametric.\citep{hu2022flexible,cheng2024doubly} The sensitivity function approach is preferred when there is insufficient knowledge about the unmeasured confounding and when the primary interest is in understanding the residual confounding in causal estimation without explicitly introducing the unmeasured confounding as random variables in causal modelling.

These two streams of sensitivity analysis for unmeasured confounding have been studied rigorously in the point treatment setting. Extensions to longitudinal causal data with time-varying treatment and time-varying unmeasured confounding are very limited and predominantly frequentist. In this paper, we proposed two Bayesian sensitivity analyses, one parametric method utilizing Bayesian g-computation based on the latent confounding variable approach \citep{keil2018bayesian} and one semi-parametric method utilizing Bayesian marginal structural models \citep{saarela2015bayesian,liu2020estimation} following the sensitivity function approach.\citep{brumback2004sensitivity} Our paper is organized as follows. Section 2 details the setup and notation. In Section 3, we provide an overview of our proposed sensitivity analysis procedure. In Section 4, we examine the performance of our proposed method using a simulation study. Section 5 demonstrates applying the proposed procedure to primary sclerosing cholangitis (PSC) registry data. A discussion of the proposed methods is provided in Section 6.

\section{Causal framework with time-varying treatment}\label{sec2}

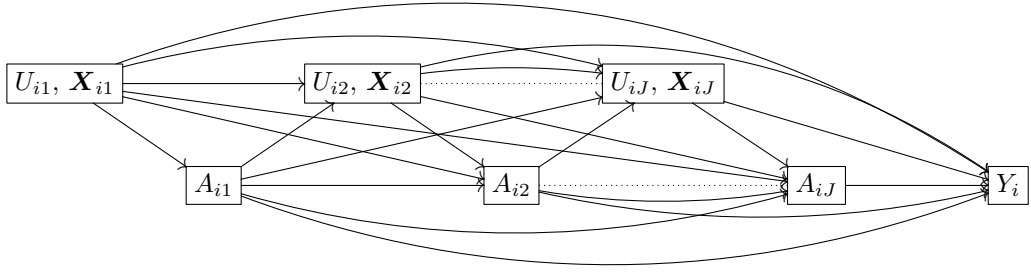
\begin{figure}[ht]
	\[ \xymatrix{ 
			&*+[o*][F]\txt{$U_{i1}$, $\boldsymbol{X}_{i1}$} \ar[dr] \ar[rr] \ar[drrr] \ar[drrrrr] \ar@/_-1.5pc/[rrrr] \ar@/_-4pc/[drrrrrrr] &
			&*+[o*][F]\txt{$U_{i2}$, $\boldsymbol{X}_{i2}$} \ar[dr] \ar[drrr] \ar@{.}[rr] \ar@/_-0.5pc/[rr] \ar@/_-2.6pc/[drrrrr]&
			&*+[o*][F]\txt{$U_{iJ}$, $\boldsymbol{X}_{iJ}$} \ar[dr] \ar[drrr]
			&\\ 
			&&*+[o*][F]\txt{$A_{i1}$} \ar[rr] \ar[ur] \ar[urrr] \ar@/_1.5pc/[rrrr] \ar@/_2.5pc/[rrrrrr]
			&&*+[o*][F]\txt{$A_{i2}$} \ar@{.}[rr] \ar[ur] \ar@/_0.5pc/[rr] \ar@/_1pc/[rrrr]
			&&*+[o*][F]\txt{$A_{iJ}$} \ar[rr]
			&&*+[o*][F]\txt{$Y_i$} 
		}
		\]
		\vspace{5mm}
	\caption{ \ \ Longitudinal causal diagram with time-varying latent variables as unobserved confounders. $U_j$, $\boldsymbol{X}_j$, and $A_j$ represent latent confounders, observed confounders, and treatment at visit $j$, respectively. $Y$ represents the end-of-study outcome.\label{DAG: method}}
\end{figure}

Suppose we have data from an observational study in which patients are treated throughout $J$ visits. We collect data of the form $\mathcal{D}_n = \{\boldsymbol X_{i1}, A_{i1},\cdots, \boldsymbol X_{iJ}, A_{iJ}, Y_i\}^n_{i=1}$, which include $n$ $i.i.d$ replicates of ($\boldsymbol X_{1}, A_{1},\cdots, \boldsymbol X_{J}, A_{J}, Y$), such that $\boldsymbol{X}_{j} \in \mathrm{R}^p$ denotes, with respectively, the measured characteristics at visit $j$; $A_j \in \{0,1\}$ denotes the treatment assignment patients received at visit $j$, $j=1, \ldots, J$; and $Y \in \mathrm{R}$ is continuous patients outcome. Let $U_{ij}$ be a time-varying unmeasured confounder for patient $i$ at visit $j$, it could be binary, categorical or continuous. Overbar notation is used to indicate the history of the variable up to visit $j$, such that $\overline{\boldsymbol X}_j = \{\boldsymbol X_1, \cdots, \boldsymbol X_j\}$ and $\overline A_j = \{ A_1, \cdots, A_j\}$. We suppress the subscript when denoting the
entire sequence, for example, $\overline{\boldsymbol X}_J = \overline {\boldsymbol X}$. The DAG that illustrates the proposed causal framework with unmeasured time-varying confounding is presented in (Figure~\ref{DAG: method}).

Following Robin's \citep{rubin1978bayesian} potential outcome framework,  define $Y(\overline a)$ as the potential outcome under the treatment sequence $\overline a$. For $J$ total visits, there are in total $2^J$ potential outcomes corresponding to $2^J$ unique binary treatment sequences $\overline a$ from set $\mathcal A = \{\overline a^1, \cdots \overline a^{2^J}\}$. We defined the causal estimand of interest as the average treatment effect (ATE), without loss of generalizability, characterized by the pair-wise mean difference between two average potential outcomes (APOs) of two distinct treatment sequences as $ATE =  E\left[Y(\overline{a})\right] -  E\left[Y(\overline{a}^*)\right]$.

Following standard causal inference assumptions with time-varying treatment, we assume stable unit treatment value assumption (SUTVA), consistency $Y(\overline a) = \sum_{\overline a \in \overline a^{2^J}}Y(\overline a)$, positivity, $0< P(A_j\mid   \overline{A}_{j-1},\overline{\boldsymbol{X}}_j, \overline{U}_{j}) < 1$ and  $0< P(A_j\mid   \overline{A}_{j-1},\overline{\boldsymbol{X}}_j) < 1$, and sequential latent ignorable treatment assumption, $Y(\overline{a})  {\perp\!\!\!\perp} A_j \mid \overline{A}_{j-1}, \overline{\boldsymbol{X}}_j, \overline{U}_{j}$. Given the complete data $\{\overline U, \overline{\boldsymbol{X}}, \overline A, Y\}$ and above assumptions, $E[Y^{\overline a}]$ is identifiable, such that
\begin{align}
E[Y(\overline{a})] & = 
\int_{x_{J}}\cdots\int_{x_{1}}\int_{u_{J}}\cdots \int_{u_{1}} E[Y(\overline{a})\mid \overline{\boldsymbol{X}} = \overline{\boldsymbol{x}},\overline{U}=\overline{u},\overline{A}_{j}=\overline{a}_{j}] \nonumber\\
& \times \prod^J_{j=1}P(U_{j}\mid \overline{U}_{j-1} = \overline{u}_{j-1}, \overline{\boldsymbol{X}}_{j}=\overline{\boldsymbol{x}}_{j},\overline{A}_{j-1}=\overline{a}_{j-1})P(\boldsymbol{X}_{j}\mid \overline{U}_{j-1}=\overline{u}_{j-1}, \overline{\boldsymbol{X}}_{j-1}=\overline{\boldsymbol{x}}_{j-1}, \overline{A}_{j-1}=\overline{a}_{j-1}) 
\label{gformula}
\end{align}
where $P(\boldsymbol{x}_{j} \mid \cdot)$ denotes the distribution of visit-specific measured confounding $\boldsymbol{x}_{j}$ and $P(U_{j} \mid \cdot)$ denotes the distribution of visit-specific unmeasured confounding $U_{j}$. 

Several causal approaches have been proposed for estimating time-varying treatment effect when $U$ is observed.\citep{robins1986new, robins2000marginal,van2006targeted,saarela2015bayesian, keil2018bayesian} The sequential latent ignorability treatment assumption does not hold without conditioning on $U$, leading to unidentifiability of the average potential outcome in (\ref{gformula}). The two proposed extended Bayesian sensitivity analysis approaches discussed in the following section can be applied to return a biased correlated causal estimand, following a plug-in type of estimation with researcher-specified plausible range of the time-varying unmeasured confounding (i.e., on the strength and direction of the relationship between the unmeasured confounding and the measured variables).

\section{Method}\label{sec3}

\subsection{Bayesian Latent Variable Sensitivity Analysis}

We proposed a Bayesian latent variable sensitivity analysis following the line of work similar to Greenland et al\citep{Greenland2003TheLeukemia} and McCandless et al.\citep{McCandless2008AConfounding} Under a Bayesian framework, one can naturally quantify the estimation uncertainty that arises from the latent confounding variable using posterior probabilistic inference and can incorporate prior evidence, expert knowledge, and external data on the distribution of the unmeasured confounding and the association between the unmeasured confounding and measured variables via prior specification. We assume $\mathcal{D}_i$ is an exchangeable sequence of real-valued random quantities over patient index with unknown parameters $\alpha$, $\gamma$, $\tau$, and $\beta$ characterizing the treatment assignment model, the observed confounder model, the latent confounder model, and the outcome model respectively. Let $\Lambda = \{ \alpha, \gamma, \tau,\beta\}$ be the set of model parameters. Specifically, we denote $\lambda = \{\alpha_U, \gamma_U,\tau_U, \beta_U\}$ as the set of bias parameters that quantify the association between the single, hypothesized time-varying unmeasured confounding $U_{ij}$ and time-varying measured variables.\citep{McCandless2017AConfounding}  $\alpha_U$ represents the association between treatment $A$ and unmeasured confounding $U$, $\gamma_U$ represents the association between measured confounding $X$ and unmeasured confounding $U$, $\tau_U$ represents the association between unmeasured confounder $U$ and other measured variables, and $\beta_U$ represents the association between outcome $Y$ and unmeasured confounder $U$. The likelihood of the observed data given parameters over $n$ subjects and $J$ visits for a binary time-varying $U$ is
\begin{align}
 \prod^n_{i=1}P(\mathcal O_i\mid \Lambda)  & = \prod ^n_{i=1} \sum_{u_{i1}=0}^1 \ldots \sum_{u_{iJ}=0}^1 P(Y_i\mid \overline{A}_{i},\overline{\boldsymbol{X}}_{i}, \overline{U}_{i},\beta) \nonumber \\
 & \times \prod^J_{j=1} \left[ P(A_{ij}\mid \overline{A}_{ij-1},\overline{\boldsymbol{X}}_{ij},\overline{U}_{ij},\alpha_j)   P(U_{ij}\mid \overline{U}_{ij-1}, \overline{A}_{ij-1}, \overline{\boldsymbol{X}}_{ij},\tau_j) P(\boldsymbol{X}_{ij}\mid \overline{\boldsymbol{X}}_{ij-1}, \overline{A}_{ij-1}, \overline{U}_{ij-1},\gamma_j) \right].\label{likelihood}
\end{align}
we assume the parameters to be a priori independent, such that $f(\Lambda) = f(\beta)f(\alpha)f(\gamma)f(\tau)$. Thus, the posterior distribution can be decomposed as $f(\Lambda\mid \mathcal \mathcal O) = f(\beta\mid \mathcal O)f(\alpha\mid \mathcal O)f(\gamma\mid \mathcal O)f(\tau\mid \mathcal O)$ (see Appendix B). Under Bayesian g-computation, APO is estimated using posterior predictive inference\citep{liu2021bayesian}, where
\begin{equation}
    \label{APO}
E\left[Y_i(\overline{a}^*)\mid \mathcal O\right] 
= \int_{\Lambda} E\left[Y_i(\overline{a}^*)\mid \Lambda\right] f(\Lambda \mid \mathcal O) d \Lambda.
\end{equation}

To carry out the sensitivity analysis, we focus on specifying the prior distributions of the bias parameters $\{\alpha_U, \gamma_U,\tau_U, \beta_U\}$, to encapsulate our assumptions and knowledge about the unmeasured confounding across visits. Bias parameters can be randomly sampled from uniform distributions \citep{McCandless2017AConfounding}. The ranges of the priors should be considered based on clinical expertise and the minimal clinically important differences. For instance, for a binary outcome, we could consider $\beta_U\sim \text{Uniform}(-2,2)$ on the log odds scale, which corresponds to an odds ratio ranging between $\exp(-2)$ and $\exp(2)$ for the effect of $U$ on the outcome $Y$. For sparse data, Gelman et al\citep{gelman2008weakly} and Ghosh et al\citep{ghosh2018use} proposed to use a default Cauchy prior with a center valued at $0$ and scale at $2.5$ or a student $t$ distribution with a degree of freedom greater than $1$ for the bias parameters.

Bayesian causal estimation can be conducted via MCMC using standard software such as JAGS and Stan.\citep{plummer2003assessment, carpenter2017stan} We will generate posterior samples of the bias and non-biased parameters and plug them into Equation \eqref{APO} to obtain the posterior predicted distribution of $E[Y_i(\overline{a})]$. Given a set of posterior draws of all parameters, $\Lambda^s$, the APO is computed in two steps for each Monte Carlo iteration $s$ as follows (suppressing covariates): 
\begin{itemize}
    \item[(1)] given $\Lambda^s$, calculate $Y_i^{s}(\overline{a})$, $i=1,\ldots, n$ over $j = 1, \ldots, J$ visits, with $E[Y_i^{s}(\overline{a})] = \int_{x_{iJ}}\cdots\int_{x_{i1}}\int_{u_{iJ}}\cdots \int_{u_{i1}} E[Y_i(\overline{a})\mid \overline{\boldsymbol{x}}_i,\overline{u}_i,\overline{A}_{ij}=\overline{a}_{ij}] \displaystyle\prod^J_{j=1}P(U_{ij}\mid \overline{u}_{ij-1}, \overline{\boldsymbol{x}}_{ij},\overline{A}_{ij-1}=\overline{a}_{ij-1})P(\boldsymbol{X}_{ij}\mid \overline{u}_{ij-1}, \overline{\boldsymbol{x}}_{ij-1}, \overline{A}_{ij-1}=\overline{a}_{ij-1}) d u_{i1}\cdots d u_{iJ}d \boldsymbol{x}_{i1}\cdots d \boldsymbol{x}_{iJ}$ 
    \item[(2)] with $Y_i^{s}(\overline{a})$, $i=1,\ldots, n$, calculate mean over $i$.
\end{itemize}
Under the full Bayesian estimation following the joint likelihood in (\ref{likelihood}), the unmeasured confounder $U$ is predicted at each Monte Carlo iteration along with posterior draws of all model parameters. However, the imputed unmeasured confounder is not directly used to estimate the APO, instead, its posterior distribution, which is conditioning on past unmeasured confounder, past treatment and measured variables, is used.  In this paper, we set the unmeasured confounders at each visit to be binary, which denotes the presence or absence of unmeasured confounding in the observed data. In practice, one can treat $U$ as a continuous, time-varying variable and proceed with the same estimation steps outlined above.

\subsection{Sensitivity function approach with time-varying unmeasured confounding}

We define the generalized sensitivity function to quantify the unmeasured time-varying confounding as follows \citep{brumback2004sensitivity,hu2022flexible,cheng2024doubly}
\begin{equation}
        c(j, \overline{a}_J, \overline{x}_J) = E[Y(\overline{a}) \mid \overline A_j = \overline a_j, \overline{X}_j=\overline{x}_j] - E[Y(\overline{a}) \mid A_j = 1 - a_j, \overline A_{j-1} = \overline a_{j-1}, \overline{X}_j=\overline{x}_j]
\label{sf_c_method}
\end{equation}
where $\overline{a}_J$ and $\overline{x}_J$ are the observed treatment history and measured confounder history for visits = $\{1, \cdots, J\}$. The sensitivity function directly captures the net difference in $Y(\overline{a})$ between those at visit $j$ treated with $a_j$ and those treated with $1 - a_j$, who have the same covariates and treatment history leading to visit $j$. When the sequential ignorability assumption holds (i.e., $Y(\overline{a}) \!\perp\!\!\!\perp A_j \mid \overline{A}_{j-1}, \overline{\boldsymbol{X}}_j$), the sensitivity function $c(j, \overline{a}_J, \overline{x}_J)$ is valued at zero indicating that there is no presence of time-varying unmeasured confounding at visit $j$. When the sequential ignorability assumption is violated in the presence of unmeasured confounding, the APO can be expressed using the sensitivity function as (see Appendix C for derivation), 
\begin{equation}
    E[Y({\overline{a}})]= E[Y \mid \overline{A} = \overline{a}] 
 - \int_{X_J}\cdots\int_{X_1} \sum_{j=1}^{J}c(j, \overline{a}_J, \overline{x}_J) P(1- a_j \mid \overline a_{j-1}, \overline x_j) P(x_j \mid \overline x_{j-1},  \overline a_{j-2}) \cdots P(x_1)  d x_1 \cdots d x_J,
     \label{sf_long}
\end{equation}
where the bias in the estimation of APO from unmeasured confounding is
\begin{equation}
    \label{sf_apobias}
        \text{Bias} = \int_{X_J}\cdots\int_{X_1} \sum_{j=1}^{J}c(j, \overline{a}_J, \overline{x}_J) P(1- a_j \mid \overline a_{j-1}, \overline x_j) P(x_j \mid \overline x_{j-1},  \overline a_{j-2}) \cdots P(x_1)  d x_1 \cdots d x_J.
\end{equation}
Therefore, the average causal effect (ATE)  comparing any two distinct treatment assignments $\overline{a}_J$ and  $\overline{a}^*_j$ is
\begin{align}
    \label{sf_ate}
     E[Y({\overline{a}_J})] - E[Y({\overline{a}^*_J})] & = E[Y \mid \overline{a}_J] - E[Y \mid \overline{a}^*_J] -  \int_{X_J}\cdots\int_{X_1} \sum_{j=1}^{J}c(j, \overline{a}_J, \overline{x}_J) P(1- a_j \mid \overline a_{j-1}, \overline x_j) P(x_j \mid \overline x_{j-1},  \overline a_{j-2}) \cdots P(x_1) \nonumber \\
 & + \sum_{j=1}^{J}c(j, \overline{a}^*_J, \overline{x}_J) P(1- a^*_j \mid \overline a^*_{j-1}, \overline x_j) P(x_j \mid \overline x_{j-1},  \overline a^*_{j-2}) \cdots P(x_1)  d x_1 \cdots d x_J. 
\end{align}

A common approach to gauge the impact of unmeasured confounding is to perform a sensitivity analysis in which the causal effect is recalculated over a grid of plausible (unmeasured) confounding strengths. While informative, this strategy merely illustrates how the estimate might vary. A more principled alternative is  to account for the bias due to unmeasured confounding directly, producing a bias-corrected causal effect estimate. \citep{Li2011PropensityConfounding,hu2022flexible}  Additionally, within a Bayesian framework, this correction naturally propagates both sampling variability and the uncertainty encoded in the sensitivity function, yielding a coherent posterior distribution for the causal effect.

\begin{equation}
      Y^{SF} = Y - \sum_{j=1}^J c(j,\overline A_J, \overline X_j)P(1-a_j\mid \overline A_{j-1}, \overline X_j).
\label{adjusted_Y}
\end{equation}

Mathematically, we can show that replacing observed outcome $Y_i$ with $Y_i^{\text{SF}}$ can effectively remove the bias in Equation \eqref{sf_ate}; details are presented in Appendix C. 
Consider the causal effect between any pair of treatments $\overline a$ and $\overline a^*$ using the adjusted outcomes $Y_i^{\text{SF}}$, the estimate of the causal effect is $ATE = E[Y^{SF}\mid \overline A_J = \overline a_J] - E[Y^{SF} \mid \overline A_J = \overline a^*_J]$. Several causal estimation approaches for time-varying treatment and confounding can be applied to estimate causal effect given the bias-corrected $Y^{SF}$, including the marginal structural models (MSM)\citep{brumback2004sensitivity} and the Bayesian marginal structural models (BMSM).\citep{saarela2015bayesian, liu2020estimation} In this study, we implemented BMSMs to provide a Bayesian estimation extension to the sensitivity function approach.   

The sensitivity function approach relies on the specification of the unidentifiable sensitivity function. Methodologies exist for devising functional forms of the sensitivity function. Interpretation of the sensitivity function for the point-treatment setting with cross-sectional data can be found in Hu et al.\citep{hu2022flexible}. We can consider the following strategies for specifying the sensitivity function $c(\cdot)$:
\begin{enumerate}
    \item Leverage subject-matter expertise to assume a plausible distribution for $c(\cdot)$, as in Brumback et al\citep{brumback2004sensitivity} and Li et al\citep{Li2011PropensityConfounding}. The $c(\cdot)$ can be expressed as a scalar parameter $\alpha$ or as a function of the measured covariates and treatment.
    \item For a binary outcome, $c(\cdot)$ denotes the expected difference between two probabilities on the outcome, with a natural bound of $[-1,1]$; For a continuous outcome, $c(\cdot)$ denotes the difference between two expected potential outcome functions. This is naturally bounded by the differences between the minimum and the maximum value of the outcome. 
    \item $c(\cdot)$ can also be specified to represent the residual confounding from the conditional outcome model that is not explained by measured covariates and treatment. \citep{hogan2014bayesian} In Hu et al, \citep{hu2022flexible} an estimated $R^2$ from the conditional outcome model was used to capture residual confounding for a binary outcome.
\end{enumerate}

\subsubsection*{Bayesian Marginal structural models with sensitivity function}

Following Saarela et al,\citep{saarela2015bayesian} the estimation of marginal treatment effect proceeds by maximizing the expected utility function (e.g., the log-likelihood of the marginal outcome model given treatment history) for a new subject following Bayesian decision theory and importance sampling. Let $u(\Theta)$ be the utility function for subject $i$. Bayesian marginal structural models maximizes the utility function with respect to the $\Theta$, the causal parameter of interest (i.e., the marginal treatment effect parameters) and the natural choice of the utility function would be the log-likelihood of the marginal treatment effect outcome model with sensitivity function corrected outcome. Let $ \mathbf{D}_n$ denote the complete observed data and $\mathcal{D}_i^*$ denote the complete observed data of a new subject $i^*$. The expectation of our utility function can be expressed using posterior predictive inference and the importance sampling technique for a new subject, such that
\begin{align}
E_{\mathcal{E}}\left[u \left(\Theta\right) \mid \mathbf{D}_n \right] 
& =\int_{\overline{d}_{i}^{*}} \log P_{\mathcal{E}}\left(y_{i}^{SF*} \mid \overline{a}_{i}^{*} ; \Theta\right) \frac{P_{\mathcal{E}}(\mathcal{D}_{i}^* \mid \mathbf{D}_{n})}{P_{\mathcal{O}}(\mathcal{D}_{i}^* \mid \mathbf{D}_{n})} P_{\mathcal{O}}\left(\mathcal{D}_{i}^{*} \mid \mathbf{D}_{n}\right) d \mathcal{D}_{i}^{*}.
\label{eq-utility}
\end{align}
Let $w_i^* = \frac{P_{\mathcal{E}}(\mathcal{D}_{i}^* \mid \mathbf{D}_n)}{P_{\mathcal{O}}(\mathcal{D}_{i}^* \mid \mathbf{D}_n)}$, which denotes the subject-specific importance sampling weight and can be dervied further as\citep{saarela2015bayesian},
\begin{align}
w_i^* &= \frac{\mathbb{E}_\alpha \left[ \prod_{j=1}^J P_{\mathcal{E}}(A_{ij}^* \mid \overline{A}_{ij-1}^*, \alpha) \mid \overline{a}_1, \ldots, \overline{a}_n) \right]}{\mathbb{E}_\beta \left[\prod_{j=1}^J  P_{\mathcal{O}}(A_{ij}^* \mid \overline{X}_{ij}^*, \overline{A}_{ij-1}^*, \beta) \mid \overline{x}_1,\ldots, \overline{x}_n,\overline{a}_1,\ldots, \overline{a}_n) \right]}.
\end{align}
We estimate the marginal treatment effect parameters $\Theta$ by maximizing (\ref{eq-utility}) with respect to $\Theta$. This is achieved in a two-step estimation process where in the first step we estimate the importance sampling weights and in the second step we use the non-parametric Bayesian Bootstrap to approximate $P_{\mathcal{O}}(\mathcal{D}_{i}^* \mid \mathbf{D}_n )$. Under the Bayesian bootstrap, $P(\pi)$, the bootstrap sampling weights are sampled from $Dir_n(1, \ldots, 1)$. The marginal treatment effect parameter $\Theta$ given the sensitvity function is then estimated via, \begin{align}
   \hat{\Theta} & = argmax_{\theta} \frac{1}{B}\sum_{b=1}^{B} \sum_{i=1}^{n} \pi_i^{(b)} \  \hat{w}_{i} \ log P_{\mathcal{E}}( y_{i} - \sum_{j=1}^{J} c(j,\overline{a}_j, \overline{x}_j) P(1-a_j \mid \overline{a}_j-1 , \overline{x}_J)) \mid \overline{a}_{i}; \Theta)
    \label{eq-maxexpectedu1-bmsms}
\end{align}  where the importance sampling treatment assignment weights, $w_{i}^*$ are estimated by the posterior mean of the predictive sequential treatment assignment density under $\mathcal{E}$ and $\mathcal{O}$ given the observed data. When $B\rightarrow \infty$, $\frac{\sum\pi_i^{(b)}}{B} \rightarrow E(\pi_i)$. A point estimate of $\Theta$ in  (\ref{eq-maxexpectedu1-bmsms}) can be returned by Monte Carlo integration over the Dirichlet $\pi_i$ draws or by approximating $\pi_i$ at its expected value $\frac{1}{n}$. The estimation uncertainty can be captured in terms of the sampling probability distribution of $\pi$ following the approximate Bayesian weighted likelihood bootstrap.\citep{newton1994approximate} We obtain a posterior distribution of $\Theta$ by repeatedly drawing $\pi$ from the uniform Dirichlet distribution and maximizing the $\pi$ weighted expected utility function. 

\section{Simulation}

\begin{figure}
\centering
\begin{subfigure}{.45\textwidth}
\centering
	\[ \xymatrix{
*++[o][F]{U_1} \ar@{->}[r] \ar@{->}[d] \ar@{->}[rrrd] & *++[o][F]{U_2} \ar@{->}[r] \ar@{->}[d] \ar@{->}[rrd] & *++[o][F]{U_3} \ar@{->}[rd] \ar@{->}[d] &  \\
*+[o*][F]{A_1} \ar@{->}[ru] \ar@{->}[r] \ar@{->}[rd] \ar@/^1pc/@{->}[rrr] & *+[o*][F]{A_2} \ar@{->}[ru] \ar@{->}[r] \ar@{->}[rd] \ar@/^1pc/@{->}[rr] &*+[o*][F]{A_3} \ar@{->}[r] & *+[o*][F]{Y} \\
*+[o*][F]{X_1} \ar@{->}[r] \ar@{->}[u] \ar@{->}[rrru] & *+[o*][F]{X_2} \ar@{->}[r] \ar@{->}[u] \ar@{->}[rru] & *+[o*][F]{X_3} \ar@{->}[ru] \ar@{->}[u] & 
}
\]		
\caption{Time-varying U\label{DAG: simulation1}}
\end{subfigure}
\begin{subfigure}{0.45\textwidth}
\centering
	\[ \xymatrix{
 & *++[o][F]{U} \ar@{->}[d] \ar@{->}[rrd] \ar@{->}[rd] \ar@{->}[ld] &  &  \\
*+[o*][F]{A_1}\ar@{->}[r] \ar@{->}[rd] \ar@/^1pc/@{->}[rrr] & *+[o*][F]{A_2} \ar@{->}[r] \ar@{->}[rd] \ar@/^1pc/@{->}[rr] & *+[o*][F]{A_3} \ar@{->}[r] & *+[o*][F]{Y} \\
*+[o*][F]{X_1} \ar@{->}[r] \ar@{->}[u] \ar@{->}[rrru] & *+[o*][F]{X_2} \ar@{->}[r] \ar@{->}[u] \ar@{->}[rru] & *+[o*][F]{X_3} \ar@{->}[ru] \ar@{->}[u] & 
}
\]
\caption{Time-invariant U\label{DAG: simulation2}}
\end{subfigure}
\caption{$\ $ \ Longitudinal causal diagram with latent variables as unobserved confounders in simulation settings.\label{DAG:simulation}}
\end{figure}
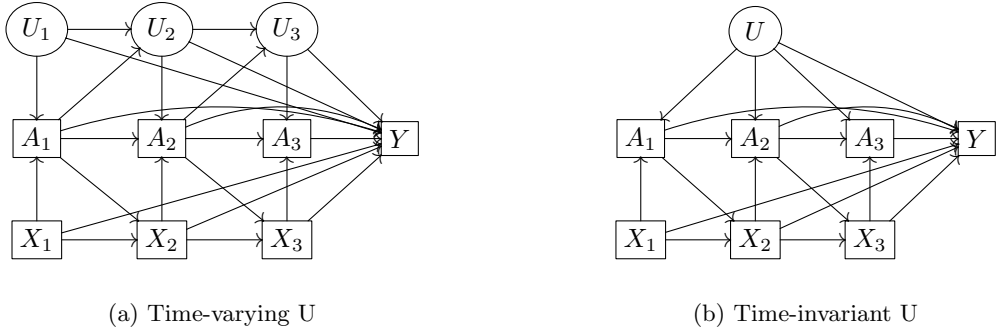

We conducted a series of simulation experiments to evaluate the effectiveness of these two proposed methods with varying sample sizes $n=500, 1000$. We considered the following simulation scenarios, time-varying unmeasured confounding (Figure~\ref{DAG:simulation}a), time-invariant unmeasured confounding (Figure~\ref{DAG:simulation}b), and no unmeasured confounding. Under time-varying unmeasured confounding, we considered three cases of unmeasured time-varying confounding: one binary unmeasured time-varying confounder, one continuous unmeasured time-varying confounder, and two time-varying unmeasured confounders (one binary and one continuous). Data-generating details were outlined in the supplementary material. 

For each simulation scenario, we performed and compared the following causal analyses: i) a naive MSM fitted to the observed data without the adjustment of unmeasured confounding (defined as the "worst-case" analysis), ii) a second MSM fitted to the complete data treating the unmeasured confounder as a measured variable (defined as the "best-case" analysis), iii) the time-varying sensitivity function approach using frequentist MSMs, iv) the time-varying sensitivity function approach using Bayesian MSMs, v) the proposed Bayesian latent variable sensitivity approach with time-invariant $U$ (BSA time-invariant $U$), and vi) the proposed Bayesian latent variable sensitivity approach with time-varying $U$ (BSA time-varying $U$). For the Bayesian latent variable approach, we assigned bias parameters except for $\sigma_U$ with prior of Uniform $(-2,2)$, which implied odds ratio for the association between $U$ and other covariates that range from $\exp(-2)= 0.14$ to $\exp(2)= 7.39$. For $\sigma_U$, we wanted to provide a wider range for the intercept, with a prior Uniform $(-5,5)$. For the sensitivity function approach, the true, simulated visit-specific sensitivity function was calculated for each subject following Equation \eqref{adjusted_Y}. We expected the sensitivity function approach to perform well, given that its true values were used in the simulation study. In practice, we can not identify the true sensitivity function given the observed data and practical guidance on how to specify the sensitivity function in an application is explained in Section 3.2 and Section 5's data analysis. To facilitate the implementation of these methods, we provided R code to replicate our simulation studies on GitHub, \url{https://github.com/reidbrok/SensitivityAnalysis_TimeVarying}.

For all simulation scenarios, the causal parameter of interest was the causal contrast in the outcome between “always treated” and “never treated”, denoted as $E(Y({\overline{1}}))-E(Y({\overline{0}}))$. We considered each simulation setting with $ns = 1000$ replications and reported the estimated mean (posterior mean for BSA) of the causal effect as $\frac{1}{ns} \sum ATE_S$, relative bias $(RB) = \frac{1}{ns}\sum_{s = 1}^{ns}\frac{\hat{ATE}_s - ATE}{ATE}$, empirical standard deviation of ATE $(ESE) = \sqrt{\frac{1}{ns}\sum(\hat{ATE}_s - ATE)^2}$, average standard error of ATE $(ASE)= \frac{1}{ns} \sum se(ATE_S)$, the relative bias of SE $(SERB) = \frac{1}{ns}\sum \frac{ASE- ESE}{ESE}$, and the 95\% coverage probability for each approach.

\begin{table}[!htbp]
\centering
\begin{tabular}{|l|l|ccccc|}
\hline
Setting & Estimator & Mean & RB & SD & SE & CP \\ \hline
\multirow{6}{*}{\begin{tabular}[c]{@{}l@{}}n = 500, Time-Varying Bernoulli $U$\\ True ATE = -10.27\end{tabular}} & MSM $U$ included & -10.28 & -0.13 & 0.66 & 0.71 & 96.30 \\ 
& MSM $U$ excluded & -11.48 & -11.76 & 0.60 & 0.63 &  49.40 \\ 
& Sensitivity Function (frequentist MSM) & -10.33 & -0.59 & 0.49 & 0.63 &   98.60\\
& Sensitivity Function (Bayesian MSM) & -10.34 &-0.64 & 0.49 & 0.57& 97.60\\ 
& BSA time-invariant $U$ & -10.58& -3.06& 0.76& 1.04& 97.10 \\ 
& BSA time-varying $U$& -9.77 & 4.90 & 0.91 & 1.58 &  99.10 \\ \hline
\multirow{6}{*}{\begin{tabular}[c]{@{}l@{}}n = 500, Time-Varying Normal $U$\\ True ATE = -9.74\end{tabular}} 
& MSM $U$ included & -9.77 & -0.35 & 0.73 & 0.59 &  93.80 \\  
& MSM $U$ excluded & -10.46 & -7.35 & 0.38 & 0.44 &  63.00 \\
& Sensitivity Function (Frequentist MSM)  &-9.69 &  0.52 & 0.32 & 0.44 & 99.40\\
& Sensitivity Function (Bayesian MSM)  &-9.69 &  0.52 & 0.31 & 0.38 & 98.10\\
& BSA time-invariant $U$ & -10.08 & -3.49 & 0.41 &0.63 & 98.10\\
& BSA time-varying $U$ & -9.59 & 1.55 & 0.48 & 1.11 &  99.90 \\  \hline
\multirow{6}{*}{\begin{tabular}[c]{@{}l@{}}n = 500, two Time-Varying $U$s,\\ one Normal $U$, one Bernoulli $U$\\ True ATE = -10.40\end{tabular}}
& MSM $U$ included & -10.43 & -0.34& 0.74 & 0.76 & 94.90 \\  
& MSM $U$ excluded & -11.84& -13.88& 0.58& 0.61&  34.80 \\
& Sensitivity Function (Frequentist MSM)  & -10.57&  -1.67 & 0.46 & 0.61  & 98.10\\
& Sensitivity Function (Bayesian MSM)  & -10.57 &  -1.70 & 0.46 & 0.55  & 97.10\\
& BSA time-invariant $U$ & -1.21 &88.17 &1.68 & 4.06 &40.00\\ 
& BSA time-varying $U$ & -10.20 &1.87& 0.91& 1.58 & 99.30 \\ \hline
\end{tabular}
\caption{\ \ Simulation results for the estimated causal parameter $E[Y(1,1,1)] - E[Y(0,0,0)]$ over 1000 replications among different time-varying confounding settings, including posterior predictive mean, relative bias (RB), standard deviation (SD), standard error (SE), and 95\%coverage probability (CP). MSM: marginal structural models; BSA: Bayesian Sensitivity Analysis approach.}
\label{simulation_result}
\end{table}

In scenarios involving time-varying unmeasured confounding as shown in Table~\ref{simulation_result}, the BSA time-varying $U$ and the two sensitivity function approaches demonstrated good performance and returned approximately unbiased causal estimators. The sensitivity function approach returned a smaller standard error as expected, where the sensitivity parameters (i.e., the visit-specific sensitivity functions) are plugged in with the true simulated values of the sensitivity function, compared to the BSA time-varying $U$ approach where we used a set of relatively vague uniform priors for the sensitvity parameters. Notably, the performance of these two approaches did not suffer when there were two unmeasured time-varying confounders. The BSA time-invariant $U$ performed well in settings with a single binary or a single continuous unmeasured time-varying confounder. However, the BSA time-invariant $U$ performed poorly when data were generated with two time-varying unmeasured confounders. This was a clear indication that when working with longitudinal causal data, one should consider implementing BSA time-varying $U$ and the sensitivity function approaches. 

\begin{figure}[!htbp]
    \centering
    \includegraphics[width=1\linewidth]{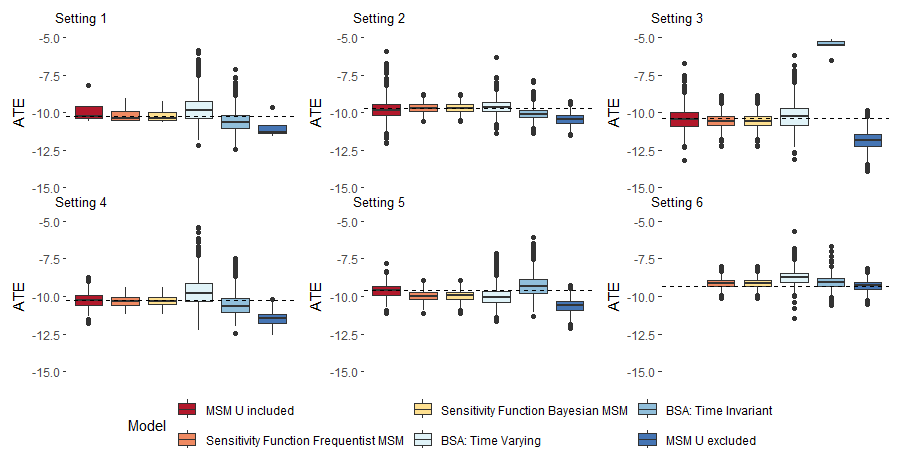}
    \caption{\ \ Estimates of causal effect ATE among 1000 replications. Setting 1: $n$ = 500 with Time-varying binary $U$; Setting 2: $n$ = 500 with Time-varying continuous $U$; Setting 3: $n$ = 500 with two unmeasured confounder $U$s, one continuous, one binary; Setting 4: $n$ = 1000 with Time-varying binary $U$;  Setting 5: $n$ = 500 with time fixed binary $U$;  Setting 6: $n$ = 500 with no presence of $U$. The ATE results that could be achieved if $U$ were observed and the naive ATE estimators excluding $U$ are also presented. Dashed lines mark the true ATE.}
    \label{fig:result}
\end{figure}

In the simulation settings where data were generated in the absent of unmeasured confounding (Setting 6 in Figure~\ref{fig:result} and Supplementary Table S1), the BSA time-invariant $U$ performed better than the BSA time-varying $U$ interms of relative bias and coverage probability (closer to nomial level). When the true data-generating mechanism involved a time-invariant binary unmeasured confounder (Setting 5 in Figure~\ref{fig:result} and Supplementary Table S1), all methods except the naive estimator (i.e., MSM excluding $U$) produced relatively unbiased point estimates. The BSA time-varying $U$ maintained good performance, suggesting flexibility even when the unmeasured confounder did not vary over time.

\section{Application to pediatric PSC Registry Data}
We applied the proposed longitudinal sensitivity analysis to a clinical dataset with time-varying treatment and confounding. The objective of the data analysis was to quantify the effectiveness of oral vancomycin therapy (OVT) for pediatric primary sclerosing cholangitis (PSC) using a multicentre PSC registry. We estimated and compared the time-varying treatment effect of OVT with and without the sensitivity analysis on unmeasured confounding. 

PSC is a chronic cholestatic liver disease characterized by ongoing inflammation, destruction, and fibrosis of intrahepatic and extrahepatic bile ducts. \citep{chapman1980primary, larusso1984current} OVT has garnered substantial interest as a potential treatment for PSC, with conflicting data on effectiveness published to date. A study conducted by Davies et al\citep{davies2008long} showed that OVT could be an effective long-term treatment of sclerosing cholangitis in children, especially those without cirrhosis. However, this conclusion is drawn from a sample size of only 14. Deneau et al\citep{deneau2021oral} utilized propensity score matching to estimate the effectiveness of OVT initiated at time of diagnosis, compared to ursodeoxycholic acid (UCDA) and no treatment. They conducted a cross-sectional causal analysis and concluded that neither OVT nor UDCA received at baseline improved biochemical or histopathological outcomes compared to no treatment. They also reported that patients who received OVT as their initial therapy were more likely to present with mild fibrosis. PSC patients can switch to or from OVT after diagnosis and the effectiveness of OVT through the course of PSC after diagnosis remains unknown. It is of great clinical interest to estimate the time-varying effect of OVT. Liu et al\citep{liu2021bayesian} conducted a Bayesian causal analysis using Bayesian latent confounding models to study the causal effect of time-varying OVT on achieving normalized Gamma-glutamyl transferase (GGT) 2 years after diagnosis. This study found a 77\% posterior predictive probability of observing a higher probability of achieving a normalized GGT at 2 years for patients who received OVT at both baseline and 1 year ("always treated") compared to patients who were OVT-free 2 years since diagnosis ("never treated"), indicating potential benefit of OVT on the early management of PSC. However, one of the limitations of these analyses was that there are potential unmeasured confounders, such as the adherence to prescribed therapies \citep{deneau2021oral, liu2021bayesian}. Here, we applied the proposed BSA and sensitivity function approaches to estimate and compare the time-varying effect of OVT in the presence of potential time-varying unmeasured confounding.

\begin{figure}[htbp]
	\[ \xymatrix{ 
			&*+[o*][F]\txt{$U_{i0}$, $X_{i0}$} \ar[dr] \ar[rr] \ar[drrr] \ar[drrrrr]  &
			&*+[o*][F]\txt{$U_{i1}$, $X_{i1}$} \ar[dr]  \ar[drrr]&
			\\ 
			&&*+[o*][F]\txt{$OVT_1$} \ar@/_1.5pc/[rrrr] \ar[rr] \ar[ru]
			&&*+[o*][F]\txt{$OVT_2$}  \ar[rr]
			&&*+[o*][F]\txt{$ln(GGT)$} 
		}
		\]
		\vspace{5mm}
  
	\caption{\ \ Longitudinal causal diagram between OVT, GGT and confounders. $U_j$, $X_j$, and $OVT$ represent latent confounders, observed confounders, and treatment at visit $j$, respectively. $ln(GGT)$ represents natural logarithm transformed GGT as the end-of-study outcome. $X_0$ and $X_1$ represent time-varying confounders including GGT, hepatobiliary events, concurrent IBD, overlapping AIH, and Metavir fibrosis stage at baseline and 1 year. In addition, $X_0$ also included time-invariant demographic confounders including age and sex. OVT: oral vancomycin;  GGT: gamma-glutamyl transferase; IBD: inflammatory bowel disease; AIH: autoimmune hepatitis.  \label{DAG: data analysis}}
\end{figure}
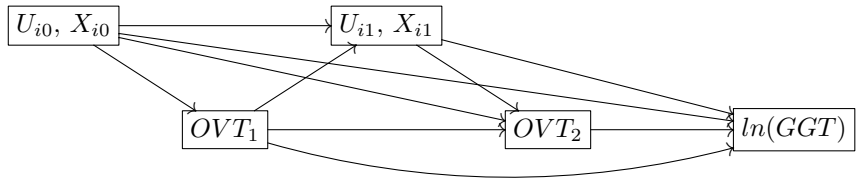

The multi-national Pediatric PSC Consortium research registry included patients diagnosed with PSC before age 18 years at any participating site. To qualify, patients had to meet PSC diagnostic criteria through cholestatic biochemistry and consistent imaging or histopathology, as outlined by Deneau et al.\citep{deneau2021oral,deneau2017natural} In what follows, we utilized only a subset of the overall registry dataset and use of the clinical data are intended solely for the purpose of illustrating methodology and not to infer any clinical conclusions. The current analysis was restricted to subjects who had complete measurements of GGT, AST, and platelet levels at the time of diagnosis and two years post-diagnosis. Exclusion criteria were patients with missing PSC onset dates and those who had received OVT before their PSC diagnosis. However, patients who underwent liver transplantation during the follow-up period were still considered for inclusion in the study. The primary outcome for the current analysis was a natural logarithm-transformed GGT at 2 years after diagnosis. Patients were defined as being exposed to OVT at each follow-up visit if they received OVT during the preceding year, based on recorded start and stop dates. There were three OVT groups: OVT-free over 2 years $(OVT = 00)$, OVT-free between 0 to 1 year and OVT exposed between 1 to 2 years $(OVT = 01)$, and OVT exposed between 0 to 1 year and remained on OVT between 1 to 2 years $(OVT = 11)$. No patients who started on OVT between 0 to 1 year stopped OVT before 2 years. We defined the cumulative treatment of OVT as the total number of follow-up intervals where patients were exposed to OVT, taking on values 0, 1, and 2. Baseline variables included age at PSC diagnosis, sex, PSC disease phenotype (large versus small duct involvement), concurrent inflammatory bowel disease (IBD), autoimmune hepatitis (AIH) overlap, Metavir fibrosis stage, hepatobiliary events and baseline GGT measure. Concurrent IBD at PSC diagnosis, at one year, and two years was defined based on whether the patient had been diagnosed with IBD before the PSC diagnosis, within one year after the PSC diagnosis, and two years after the PSC diagnosis, respectively. Similarly, AIH overlap was determined if a patient had a recorded diagnosis of AIH before the PSC diagnosis, within one year after the PSC diagnosis, and within two years after the PSC diagnosis. The Metavir fibrosis stage at each visit was defined as the stage recorded from the closest documented biopsy within one year prior to that visit, with the possibility of patients undergoing multiple liver biopsies during the follow-up period. Whether or not a hepatobiliary event had occurred was assessed at baseline and one year based on event occurrence date. GGT was measured at annual intervals following PSC diagnosis. The proposed causal diagram is provided in Figure \ref{DAG: data analysis}. 

We conducted three causal analyses on the complete data analysis: the conventional marginal structure models without consideration of unmeasured confounding, and the BSA and sensitivity function with time-varying unmeasured confounding. We were interested in estimating the changes in the mean of natural logarithm-transformed GGT at 2 years after diagnosis between patients who were exposed to OVT between 0 to 1 year and between 1 to 2 years (OVT = 11, “always treated”) and patients who were OVT free over 2 years (OVT = 00, “never treated”). 

For the MSM model, the visit-specific conditional treatment assignment model was fitted via logistic regression, adjusting for OVT exposure at the previous visit, hepatobiliary events, clinical measurements (concurrent IBD, overlapping AIH and Metavir fibrosis stages) and prescription of UDCA and concomitant medications at the current visit, and patient’s age, sex and duct phenotype at baseline. 

In the Bayesian latent variable approach, we constructed a set of models to capture the relationship between time-varying unmeasured confounder $U$ and other observed variables ($ln(GGT), OVT, \boldsymbol{X}$). The coefficients that represent the association between unmeasured confounder $U$ and other observed variables were labelled as bias parameters. When the values of these sensitivity parameters are equal to zero, it implies the absence of an unmeasured confounder. We specified the following models:
\begin{flalign*}
&\boldsymbol{X_0} \sim Bernoulli (\boldsymbol\gamma_{00})& \\
&U_{0} \mid \boldsymbol X_0 \sim Bernoulli(\tau_{00})& \\
&OVT_1 \mid  \boldsymbol X_0, U_0 \sim Bernoulli(\alpha_{00} + \alpha_{01}\boldsymbol X_0 + \alpha_{02} U_0)&\\
&\boldsymbol X_{1} \mid \boldsymbol X_0, OVT_1, U_0 \sim Bernoulli(\gamma_{10} +   \gamma_{1}  OVT_1 +   \gamma_2  U_0 + \gamma_3  \boldsymbol X_0 )&\\
&U_{1} \mid \boldsymbol X_1, OVT_1, U_0, \boldsymbol X_0 \sim Bernoulli(\tau_{10} +   \tau_1  OVT_1 +   \tau_2  U_0 + \tau_3  \boldsymbol X_1 + \tau_4 \boldsymbol X_0)\\
&OVT_2 \mid  \overline{\boldsymbol{X}}, OVT_1, \overline{U}  \sim Bernoulli(\alpha_{10} +   \alpha_{11}  OVT_1 +   \alpha_{12}  U_0 + \alpha_{13}  U_1 + \alpha_{14}  \boldsymbol X_1 + \alpha_{15}\boldsymbol X_0)&\\
&ln(GGT) \mid  \overline{\boldsymbol{X}}, \overline{OVT}, \overline{U} \sim Normal(\beta_0 +{\beta_1} \boldsymbol X_0 + {\beta_2} \boldsymbol X_1 +\beta_3 OVT_1 +\beta_4 OVT_2 + \beta_5 U_0 + \beta_6 U_1, \tau_{sd})& 
\end{flalign*}
The bias parameters, \{$\gamma_2, \alpha_{02}, \alpha_{12}, \alpha_{13}, \beta_{5}, \beta_{6}$\}, presented the association between the time-varying unmeasured confounding and measured variables. Vague uniform priors ranging from -10 to 10 were used for all bias parameters, and normal priors with zero means were assigned for other non-bias parameters and the precision was 0.01 and 0.001 for coefficients and intercepts, respectively. We saved 15,000 MCMC draws using 25,000 burn-in iterations and 25,000 further iterations with every fifth sample being gathered. We assumed the unmeasured time-varying confounder $U$ is dependent on the measured time-varying confounder $X$ at current and previous visits. MCMC convergence was evaluated using Geweke’s Z-score and an absolute value less than 2 was accepted. \citep{geweke1991evaluating} All statistical analyses were conducted using R software (version 4.3.2) and Just Another Gibbs Sampler (JAGS, version 4.3.0).\citep{plummer2003assessment}

For the implementation of the Bayesian sensitivity function approach, we defined the sensitivity function as 
\begin{equation}
    \label{da: sf_c}
    \begin{aligned}
        c(j, \overline{a}_J, \boldsymbol{\overline{x}}_J) = E[ln(GGT)(\overline{a})\mid \overline{OVT} = \overline{a}_j, \boldsymbol{\overline{X}}_j = \boldsymbol{\overline{x}}_j] - E[ln(GGT)({\overline{a}})\mid {OVT}_j= 1- a_j, \overline{OVT}_{j-1} = \overline a_{j-1} \boldsymbol{\overline{X}}_j = \boldsymbol{\overline{x}}_j]
    \end{aligned}
\end{equation}
where $\overline{a}$ represents the observed treatment assignment history across 2 years of measurement. Since there is no sufficient domain knowledge to inform the unmeasured confounding, we adopt the specification of the sensitivity function using the residual standard deviation estimated from the conditional outcome model, $\hat{\sigma}$, similar to the sensitivity function specification discussed in the data application in Hu et al.\citep{hu2022flexible}. We assumed the bias falls within the plausible range of $h\hat{\sigma}$, centring around the estimated sensitivity function $\pm \hat{\sigma}$, where $h=1$. 

The final cohort included 401 PSC patients diagnosed between July 1988 and December 2017. Of these, 343 patients were OVT-free over the first 2 years after diagnosis, 19 patients were prescribed OVT only between 1 to 2 years after diagnosis and 39 patients were prescribed OVT between 0 to 1 year and remained on OVT at 2 years. The baseline characteristics of each exposure group were shown in the Supplementary Table S2. 

The estimated cumulative treatment effect of OVT on the end-of-study outcome, natural logarithm transformed $ln(GGT)$, including results from the sensitivity analyses, were shown in Table \ref{da:result}. The conventional marginal structure model that ignored the unmeasured confounding reported an average of -0.33 unit (95\% Confidence Interval:-0.80, 0.14) change in $ln(GGT)$.  The time-varying BSA returned an average of -0.29 units (95\% Credible Interval: -0.73,0.16) change in $ln(GGT)$, and the Bayesian sensitivity function with $h=1$ returned an average of -0.34 units (95\% Credible Interval: -0.74, 0.12) change in $ln(GGT)$, respectively. Both sensitivity analyses revealed minimal impact of the potential unmeasured time-varying confounding on the causal estimation.

\begin{table}[!htbp]
\centering
\caption{\ \ Estimated treatment effect of OVT on study outcomes. The columns are mean, standard deviation, and 95\% CI of the posterior predictive distribution of the causal effect. $ln(GGT)$: natural logarithm transformed gamma-glutamyl transferase; OVT: oral vancomycin therapy; MSM: marginal structure model; BSA: Bayesian sensitivity analysis. CI: credible interval or confidence interval \label{da:result}}
\begin{tabular}{llll}
\hline\hline
\multicolumn{1}{l}{}    & \multicolumn{1}{l}{\textbf{Mean}} & \multicolumn{1}{l}{\textbf{SD}}                         & \textbf{95\% CI} \\
\multicolumn{1}{l}{OVT on end-study measured $ln(GGT)$: 11 vs 00 }                                                                  \\ \hline
MSM & -0.33 & 0.24 & (-0.80, 0.14)\\
Sensitivity Function with Bayesian MSM & -0.34 & 0.33 & (-0.74, 0.12)\\
BSA & -0.29 & 0.23 & (-0.73, 0.16)\\ \hline
\end{tabular}
\end{table}

\section{Discussion}

In this study, we introduced two Bayesian approaches for causal sensitivity analysis in the presence of unmeasured time-varying confounding: the Bayesian latent variable approach and the Bayesian sensitivity function approach. The latent variable approach models time-varying unmeasured confounding as hidden variables within the causal model, enabling the integration of external data or prior knowledge and offering greater conceptual and interpretative clarity in applied settings. In contrast, the sensitivity function approach directly characterizes the net bias induced by unmeasured time-varying confounding without explicitly modelling latent variables, thereby avoiding distributional assumptions; however, this approach may be less interpretable in practice. To the best of our knowledge, this is the first study to formalize the derivation of both approaches and to provide a comprehensive comparison through simulation.

The Bayesian latent variable approach incorporates treatment, outcome, and covariate model-specific bias parameters to quantify the relationship between unobserved and observed confounders. This is in line with other probabilistic sensitivity analysis methods that leverage prior knowledge of the bias parameters to conceptualize the impact of time-varying unmeasured confounding, which naturally captures estimation uncertainty. The proposed Bayesian latent variable approach can be easily implemented in R statistical software and off-the-self MCMC software such as JAGS and STAN. \citep{plummer2003assessment, carpenter2017stan} The Bayesian sensitivity function approach with Bayesian marginal structural models has a natural connection to the frequentist sensitivity function approach with frequentist marginal structural models. Bayesian causal estimation under this method can be performed semi-parametrically or non-parametrically, following maximization of the weighted utility function via Bayesian non-parametric bootstrap. \citep{saarela2015bayesian,stephens2023causal,luo2025longitudinal}

Our simulation study reveals that the proposed Bayesian latent variable approach with a time-varying latent confounder offers a practical strategy for addressing unmeasured confounding in longitudinal settings and can be extremely useful when there is historical or external data on the unmeasured confounding. While the sensitivity function methods can yield more efficient and unbiased estimates, they rely on the correct specification of the sensitivity function and are difficult to interpret and explain clinically. We recommend in practice to conduct both the Bayesian latent variable approach and the Bayesian sensitivity function approach, and to discuss the sensitivity analysis results with clinical collaborators to ensure sound and interpretable conclusions. Caution is warranted in settings where the unmeasured confounding structure is complex, such as when multiple confounders with differing temporal patterns are present. In such cases, the latent variable formulation underlying the BSA approach may be misspecified, potentially limiting its validity, as reflected in our simulation results, where the Bayesian latent variable approach with a time-invariant latent variable performs poorly when the simulated causal data contains two time-varying unmeasured confounders.

The Bayesian latent variable approach relied on the correct parametric model specification of the treatment, outcome, measured and unmeasured confounding variables, while the Bayesian sensitivity function approach relied on the correct ``guess" of the sensitivity function. Our simulation study did not consider scenarios of a mis-specificed conditional outcome model (e.g., the true outcome model features non-linear terms where the outcome model in BSA is specified linearly) and scenarios where there is a skewed time-varying unmeasured confounder in the simulated data. These more complicated causal structures can occur in data applications. Future work will explore this limitation by extending these two Bayesian causal sensitivity approaches with flexible estimation algorithms, such as Bayesian additive regression tree.\citep{hu2022flexible} While this study focuses on a binary treatment and a continuous or binary end-of-study outcome, future research will explore more complex longitudinal causal structures. These include scenarios with repeatedly measured outcomes, multiple treatments, and time-to-event outcomes. 

In conclusion, this study presents a Bayesian sensitivity analysis framework that effectively incorporates uncertainty surrounding bias parameters to provide bias-corrected causal effect estimates. Our work contributes to the broader goal of improving the reliability and validity of observational studies in clinical and epidemiological research. Future research will build on these foundations, extending the methodology to more complex longitudinal causal structures.

\section*{Acknowledgements}
Dr. Liu is supported by the Connaught Fund New Investigator Award and the University of Toronto Data Sciences Institute Seed Funding for Methodologists. The reported PSC analysis and data were supported by PSC Partners Seeking A Cure, the Primary Children’s Hospital Foundation, and the National Center for Advancing Translational Sciences of the National Institutes of Health under Award Numbers KL2TR001065 and 8UL1TR000105 (formerly UL1RR025764). The content is solely the responsibility of the authors and does not necessarily represent the official views of the National Institutes of Health.

\section*{Data Availability Statement}

Access to data from the Pediatric PSC Consortium may be available upon request to the registry sponsor.

\bibliographystyle{chicago}
\bibliography{reference}

\section*{Appendix A. Identifiability with unmeasured confounding }\label{Append.A}

\begin{align}
E[Y_i(\overline{a})] & = \int_{\boldsymbol{x}_{i1}}\int_{u_{i1}}E[Y_i(\overline{a})\mid  \boldsymbol{x}_{i1},u_{i1}]P(u_{i1}\mid \boldsymbol{x}_{i1})P(x_{i})d u_{i1}d x_{i1} \nonumber \\
& =\int_{\boldsymbol{x}_{i1}}\int_{u_{i1}}E[Y_i(\overline{a})\mid A_{i1}=a_{i1},\boldsymbol{x}_{i1},u_{i1}]P(u_{i1}\mid \boldsymbol{x}_{i1})P(x_{i1}
)d u_{i1}d \boldsymbol{x}_{i1} \nonumber \\
&=\int_{\boldsymbol{x}_{i1}}\int_{\boldsymbol{x}_{i2}}\int_{u_{i1}}\int_{u_{i2}}E[Y_i(\overline{a})\mid A_{i1}=a_{i1},\boldsymbol{x}_{i1},u_{i1},\boldsymbol{x}_{i2},u_{i2}, \theta]P(u_{i2}\mid A_{i1}=a_{i1}, u_{i1},\boldsymbol{x}_{i2},\boldsymbol{x}_{i1}) \nonumber \\ 
& \times P(\boldsymbol{x}_{i2}\mid A_{i1}=a_{i1}, u_{i1},\boldsymbol{x}_{i1})P(u_{i1}\mid x_{i1})P(\boldsymbol{x}_{i1})d u_{i2}d u_{i1}d \boldsymbol{x}_{i2}d x_{i1} \nonumber \\
& =  \int_{\boldsymbol{x}_{iJ}}\cdots \int_{\boldsymbol{x}_{i1}}\int_{u_{iJ}}\cdots \int_{u_{i1}}E[Y_i(\overline{a})\mid \overline{A}_{ij}=\overline{a}_{ij}, \overline{\boldsymbol{X}}_i, \overline{U}_i, ]P(U_{ij}\mid \overline{U}_{ij-1},\overline{\boldsymbol{X}}_{ij},\overline{A}_{ij-1}=\overline{a}_{ij-1})   \nonumber \\ 
& \times P(\boldsymbol{X}_{ij}\mid \overline{U}_{ij-1}, \overline{\boldsymbol{X}}_{ij-1}, \overline{A}_{ij-1}=\overline{a}_{ij-1})\cdots P(U_{i1}\mid \boldsymbol{X}_{i1})P(\boldsymbol{X}_{i1})d u_{i1}\cdots d u_{iJ}d \boldsymbol{x}_{i1}\cdots d \boldsymbol{x}_{iJ}  \nonumber \\
&= \int_{x_{iJ}}\cdots\int_{x_{i1}}\int_{u_{iJ}}\cdots \int_{u_{i1}} E[Y_i(\overline{a})\mid \overline{\boldsymbol{x}}_i,\overline{u}_i,\overline{A}_{ij}=\overline{a}_{ij}]\prod^J_{j=2}P(U_{ij}\mid \overline{u}_{ij-1},\overline{\boldsymbol{x}}_{ij},\overline{A}_{ij-1}=\overline{a}_{ij-1}) \nonumber \\
& \times P(\boldsymbol{X}_{ij}\mid \overline{u}_{ij-1}, \overline{\boldsymbol{x}}_{ij-1}, \overline{A}_{ij-1}=\overline{a}_{ij-1})P(U_{i1}\mid \boldsymbol{x}_{i1})P(\boldsymbol{X}_{i1}) d u_{i1}\cdots d u_{iJ}d \boldsymbol{x}_{i1}\cdots d \boldsymbol{x}_{iJ}  
\end{align}
where $P(\boldsymbol{X}_{ij} \mid \cdot)$ denotes the distribution of visit-specific covariates $\boldsymbol{X}_{ij}$ and $P(U_{ij} \mid \cdot)$ denotes the distribution of visit-specific unmeasured confounding $U_{ij}$. 

\section*{Appendix B. Posterior distribution of the Bayesian sensitivity analysis}\label{Append.B}
\begin{align}
\label{posterior distribution}
f(\Lambda\mid \mathcal O) 
&= \frac{\prod^J_{j=1}\prod ^n_{i=1} \sum_{u_{i1}=0}^1 \ldots \sum_{u_{iJ}=0}^1 P(Y_i\mid \overline{A}_{i},\overline{\boldsymbol{X}}_{i}, \overline{U}_{i},\beta)f(\beta)}{\int_\beta\prod ^n_{i=1} \sum_{u_{i1}=0}^1 \ldots \sum_{u_{iJ}=0}^1 P(Y_i\mid \overline{A}_{i},\overline{\boldsymbol{X}}_{i}, \overline{U}_{i},\beta) f(\beta)d \beta} \nonumber \\
& \times \hspace{16pt} \frac{\prod ^n_{i=1} \sum_{u_{i1}=0}^1 \ldots \sum_{u_{iJ}=0}^1 P(A_{ij}\mid \overline{A}_{ij-1},\overline{\boldsymbol{X}}_{ij},\overline{U}_{ij},\alpha) f(\alpha)}{\int_\alpha \prod^J_{j=1}\prod ^n_{i=1} \sum_{u_{i1}=0}^1 \ldots \sum_{u_{iJ}=0}^1P(A_{ij}\mid \overline{A}_{ij-1},\overline{\boldsymbol{X}}_{ij},\overline{U}_{ij},\alpha) f(\alpha)d \alpha} \nonumber \\
& \times \hspace{16pt} \frac{\prod ^n_{i=1} \sum_{u_{i1}=0}^1 \ldots \sum_{u_{iJ}=0}^1 P(U_{ij}\mid \overline{U}_{ij-1}, \overline{A}_{ij-1}, \overline{\boldsymbol{X}}_{ij},\tau) f(\tau)}{\int_\tau \prod^J_{j=1}\prod ^n_{i=1} \sum_{u_{i1}=0}^1 \ldots \sum_{u_{iJ}=0}^1 P(U_{ij}\mid \overline{U}_{ij-1}, \overline{A}_{ij-1}, \overline{\boldsymbol{X}}_{ij},\tau) f(\tau)d \tau} \nonumber \\
& \times \hspace{16pt} \frac{\prod ^n_{i=1} \sum_{u_{i1}=0}^1 \ldots \sum_{u_{iJ}=0}^1 P(\boldsymbol{X}_{ij}\mid \overline{\boldsymbol{X}}_{ij-1}, \overline{A}_{ij-1}, \overline{U}_{ij-1},\gamma) f(\gamma)}{\int_\gamma \prod^J_{j=1}\prod ^n_{i=1} \sum_{u_{i1}=0}^1 \ldots \sum_{u_{iJ}=0}^1P(\boldsymbol{X}_{ij}\mid \overline{\boldsymbol{X}}_{ij-1}, \overline{A}_{ij-1}, \overline{U}_{ij-1},\gamma) f(\gamma)d \gamma} \nonumber  \\
& =  f(\beta\mid \mathcal O)f(\alpha\mid \mathcal O)f(\gamma\mid \mathcal O)f(\tau\mid \mathcal O)
\end{align}

\section*{Appendix C. Sensitivity function derivation}\label{Append.C}
With sensitivity function defined as 
\begin{equation}
\label{sf}
    c(j, \overline{a}_J, \overline{x}_J) = E[Y(\overline{a}) \mid \overline A_j = \overline a_j, \overline{x}_j] - E[Y(\overline{a}) \mid A_j = 1 - a_j, \overline A_{j-1} = \overline a_{j-1}, \overline{x}_j]
\end{equation}
where $\overline{a}_J$ and $\overline{x}_J$ are the observed treatment history and measured confounder history for visits = $\{1, \cdots, J\}$.
\begin{align}
    \label{sf_apo}
     E[Y(\overline{a})] &= \int_{x_1} \sum_{A_1} E[Y(\overline{a}) \mid A_1, x_1] P(A_1 \mid x_1) P(x_1) d x_1 \nonumber \\
    &=  \int_{x_1} E[Y(\overline{a}) \mid A_1 = a_1, x_1] P(a_1 \mid x_1) P(x_1) + E[Y(\overline{a}) \mid A_1 = 1- a_1, x_1] P(1- a_1 \mid x_1) P(x_1)  d x_1 \nonumber \\
 &= \int_{x_2}\int_{x_1} \sum_{A_2}\sum_{A_1}E[Y(\overline{a}) \mid \overline{A}_2, \overline x_2] P(A_2 \mid A_1, \overline x_2) P(x_2 \mid A_1, x_1)P(A_1 \mid x_1) P(x_1) d x_1 d x_2\nonumber \\
 &= \int_{x_2}\int_{x_1} E[Y(\overline{a}) \mid \overline{A}_2 = \overline a_2, \overline x_2] P(A_2 =a_2\mid A_1 = a_1, \overline x_2) P(x_2 \mid A_1=a_1, x_1)P(A_1=a_1 \mid x_1) P(x_1) \nonumber \\
 & = \int_{x_J}\cdots\int_{x_1} \sum_{A_J}\cdots\sum_{A_1} E[Y(\overline{a}) \mid \overline{A}_J, \overline x_J] P(A_J \mid \overline A_{J-1}, \overline x_{J-1}) P(x_{J-1} \mid \overline A_{J-1}, \overline x_{J-2})\cdots P(A_1 \mid x_1) P(x_1) d x_1 \cdots d x_J\nonumber \\
 &= E[Y \mid \overline{A}_J = \overline{a}_J] 
 - \int_{x_J}\cdots\int_{x_1} \sum_{j=1}^{J}c(j, \overline{a}_J, \overline{x}_J) P(1- a_j \mid \overline a_{j-1}, \overline x_j) P(x_j \mid \overline x_{j-1},  \overline a_{j-2}) \cdots P(x_1)  d x_1 \cdots d x_J
\end{align}

We can use $Y^{SF}_i$ that removes the confounding bias following Equation~\eqref{sf_apo}, 
where 

\begin{equation}
      Y^{SF} = Y - \sum_{j=1}^J c(j,\overline A_J, \overline X_j)P(1-a_j\mid \overline A_{j-1}, \overline X_j)  
\end{equation}

The estimation of the average potential outcome based on the adjusted outcome $Y^{SF}$ removes the bias from \eqref{sf_apo} 

\begin{align}
    & E[Y \mid \overline{A}_J = \overline{a}_J] 
 - \int_{x_J}\cdots\int_{x_1} \sum_{j=1}^{J}c(j, \overline{a}_J, \overline{x}_J) P(1- a_j \mid \overline a_{j-1}, \overline x_j) P(x_j \mid \overline x_{j-1},  \overline a_{j-2}) \cdots P(x_1)  d x_1 \cdots d x_J \nonumber \\
 =& \int_{x_J}\cdots\int_{x_1} \left(E[Y \mid \overline{A}_J = \overline{a}_J, \overline{X}_J = \overline{x}_J] 
 - \sum_{j=1}^{J}c(j, \overline{a}_J, \overline{x}_J) P(1- a_j \mid \overline a_{j-1}, \overline x_j) \right) P(x_J \mid \overline x_{J-1},  \overline a_{J-2}) \cdots P(x_1)  d x_1 \cdots d x_J \nonumber \\
 = & \int_{x_J}\cdots\int_{x_1} E[Y- \sum_{j=1}^{J}c(j, \overline{a}_J, \overline{x}_J) P(1- a_j \mid \overline a_{j-1}, \overline x_j) \mid \overline{A}_J = \overline{a}_J, \overline{X}_J = \overline{x}_J] P(x_J \mid \overline x_{J-1},  \overline a_{J-2}) \cdots P(x_1)  d x_1 \cdots d x_J \nonumber \\
  = & \int_{x_J}\cdots\int_{x_1} E[Y^{SF} \mid \overline{A}_J = \overline{a}_J, \overline{X}_J = \overline{x}_J] P(x_J \mid \overline x_{J-1},  \overline a_{J-2}) \cdots P(x_1)  d x_1 \cdots d x_J
\end{align}
If the sensitivity function for any distinct pair of treatment sequences is zero then the APO formulation \eqref{sf_apo} is simplified to the conventional g-formula with no unmeasured confounding. We can write out the casual effect on any pairwise average treatment effect as, 

\begin{align}
    \label{sf_ate2}
        & ATE  = E[Y({\overline{a}_J})] - E[Y({\overline{a}^*_J})] \nonumber \\
        & = E[Y \mid \overline{a}_J] 
 - \int_{x_J}\cdots\int_{x_1} \sum_{j=1}^{J}c(j, \overline{a}_J, \overline{x}_J) P(1- a_j \mid \overline a_{j-1}, \overline x_j) P(x_j \mid \overline x_{j-1},  \overline a_{j-2}) \cdots P(x_1)  d x_1 \cdots d x_J \nonumber \\
 & - E[Y \mid \overline{a}^*_J] 
 - \int_{x_J}\cdots\int_{x_1} \sum_{j=1}^{J}c(j, \overline{a}^*_J, \overline{x}_J) P(1- a^*_j \mid \overline a^*_{j-1}, \overline x_j) P(x_j \mid \overline x_{j-1},  \overline a^*_{j-2}) \cdots P(x_1)  d x_1 \cdots d x_J \nonumber \\
 & = E[Y \mid \overline{a}_J] - E[Y \mid \overline{a}^*_J] -  \int_{x_J}\cdots\int_{x_1} \sum_{j=1}^{J}c(j, \overline{a}_J, \overline{x}_J) P(1- a_j \mid \overline a_{j-1}, \overline x_j) P(x_j \mid \overline x_{j-1},  \overline a_{j-2}) \cdots P(x_1) \nonumber \\
 & + \sum_{j=1}^{J}c(j, \overline{a}^*_J, \overline{x}_J) P(1- a^*_j \mid \overline a^*_{j-1}, \overline x_j) P(x_j \mid \overline x_{j-1},  \overline a^*_{j-2}) \cdots P(x_1)  d x_1 \cdots d x_J 
\end{align}

\section*{Supplementary Materials}
\subsection*{1. Simulation set-up}

\subsubsection*{Time-varying binary unmeasured confounder}

We simulated a data set that contains three visits, with $n$ subjects. For each subject $i$, at visit $j$, $j=1, 2, 3$, we simulated one binary time-varying unmeasured confounder $U_{ij}$, one binary time-varying measured confounder $X_{ij}$, and a binary time-varying treatment $A_{ij}$, and a continuous outcome $Y_{i}$. 
\begin{align*}
U_{1} &\sim \text{Bernoulli}(0.5) \\
X_{1} &\sim \text{Bernoulli}(0.5) \\
P(A_{1}=1 \mid U_{1},X_{1}) 
  &= \operatorname{logit}^{-1}\!\bigl(1 + 0.5 X_{1} - 0.5 U_{1}\bigr) \\
P(U_{2}=1 \mid U_{1},A_{1}) 
  &= \operatorname{logit}^{-1}\!\bigl(-0.5 U_{1} - 0.5 A_{1}\bigr) \\
P(X_{2}=1 \mid X_{1},A_{1}) 
  &= \operatorname{logit}^{-1}\!\bigl(0.5 X_{1} + 0.5 A_{1}\bigr) \\
P(A_{2}=1 \mid U_{1},U_{2},X_{1},X_{2},A_{1}) 
  &= \operatorname{logit}^{-1}\!\bigl(1 - 0.3 A_{1} 
       + 0.4 X_{1} + 0.5 X_{2} 
       - 0.4 U_{1} - 0.5 U_{2}\bigr) \\
P(U_{3}=1 \mid U_{2},A_{2}) 
  &= \operatorname{logit}^{-1}\!\bigl(-0.5 U_{2} - 0.5 A_{2}\bigr) \\
P(X_{3}=1 \mid X_{2},A_{2}) 
  &= \operatorname{logit}^{-1}\!\bigl(0.5 X_{2} + 0.5 A_{2}\bigr) \\
P(A_{3}=1 \mid U_{1},U_{2},U_{3},X_{1},X_{2},X_{3},A_{2}) 
  &= \operatorname{logit}^{-1}\!\bigl(1 - 0.3 A_{2} 
       + 0.4 X_{1} + 0.5 X_{2} + 0.6 X_{3} 
       - 0.3 U_{1} - 0.4 U_{2} - 0.5 U_{3}\bigr) \\
Y \mid \overline{A}_{3},\overline{\boldsymbol X}_{3},\overline{U}_{3} 
  &\sim \text{Normal} \ \!\Bigl(
       10 - 4 A_{3} - 3 A_{2} - 2 A_{1}
       -   X_{1} - 2 X_{2} - 3 X_{3} 
       +   U_{1} + 2 U_{2} + 3 U_{3},
       2\Bigr)
\end{align*}

\subsubsection{Time-invariant binary unmeasured confounder}

We simulated another set of longitudinal data with $n$ subjects and three visits, and a time-invariant unmeasured confounder. For each subject $i$, at visit $j$, $j=1, 2, 3$, we simulated a binary time-invariant unmeasured confounder $U$, binary time-varying measured confounder $X_{ij}$, a binary time-varying treatment $A_{ij}$, and a continuous outcome $Y_{i}$. 
\begin{align*}
U      &\sim \text{Bernoulli}(0.4) \\
X_{1}  &\sim \text{Bernoulli}(0.5) \\
P(A_{1}=1 \mid U,X_{1}) 
  &= \operatorname{logit}^{-1}\!\bigl(0.5 X_{1} - 0.5 U\bigr) \\
P(X_{2}=1 \mid X_{1},A_{1}) 
  &= \operatorname{logit}^{-1}\!\bigl(0.5 X_{1} + 0.5 A_{1}\bigr) \\
P(A_{2}=1 \mid U,X_{1},X_{2},A_{1}) 
  &= \operatorname{logit}^{-1}\!\bigl(0.5 A_{1} + 0.5 X_{1} + 0.5 X_{2} - 0.5 U\bigr) \\
P(X_{3}=1 \mid X_{2},A_{2}) 
  &= \operatorname{logit}^{-1}\!\bigl(0.5 X_{2} + 0.5 A_{2}\bigr) \\
P(A_{3}=1 \mid U,X_{1},X_{2},X_{3},A_{2}) 
  &= \operatorname{logit}^{-1}\!\bigl(0.5 A_{2} + 0.5 X_{1} + 0.5 X_{2} + 0.5 X_{3} - 0.5 U\bigr) \\
Y \mid \overline{A}_{3},\overline{\boldsymbol X}_{3},U 
  &\sim \text{Normal} \ \!\Bigl(
       10 - 4 A_{3} - 3 A_{2} - 2 A_{1}
       -   X_{1} - 2 X_{2} - 3 X_{3}
       + 3\,U,\,
       2\Bigr)
\end{align*}

\subsubsection{Time-varying continuous unmeasured confounder}
For each subject $i$, at visit $j$, $j=1, 2, 3$, we simulated one continuous time-varying unmeasured confounder $U_{ij}$, one binary time-varying measured confounder $X_{ij}$, and a binary time-varying treatment $A_{ij}$, and a continuous outcome $Y_{i}$. 
\begin{align*}
U_{1} &\sim \text{Normal}(0.5,\,2) \\
X_{1} &\sim \text{Bernoulli}(0.5) \\
P(A_{1}=1 \mid U_{1},X_{1}) 
  &= \operatorname{logit}^{-1}\!\bigl(0.5 X_{1} - 0.5 U_{1}\bigr) \\
U_{2} \mid U_{1},A_{1} 
  &\sim \text{Normal}\!\bigl(-0.5 U_{1} - 0.5 A_{1},\,2\bigr) \\
P(X_{2}=1 \mid X_{1},A_{1}) 
  &= \operatorname{logit}^{-1}\!\bigl(0.5 X_{1} + 0.5 A_{1}\bigr) \\
P(A_{2}=1 \mid U_{1},U_{2},X_{1},X_{2},A_{1}) 
  &= \operatorname{logit}^{-1}\!\bigl(-0.3 A_{1} 
       + 0.4 X_{1} + 0.5 X_{2} 
       - 0.2 U_{1} - 0.2 U_{2}\bigr) \\
U_{3} \mid U_{2},A_{2} 
  &\sim \text{Normal}\!\bigl(-0.5 U_{2} - 0.5 A_{2},\,2\bigr) \\
P(X_{3}=1 \mid X_{2},A_{2}) 
  &= \operatorname{logit}^{-1}\!\bigl(0.5 X_{2} + 0.5 A_{2}\bigr) \\
P(A_{3}=1 \mid U_{1},U_{2},U_{3},X_{1},X_{2},X_{3},A_{2}) 
  &= \operatorname{logit}^{-1}\!\bigl(-0.3 A_{2} 
       + 0.4 X_{1} + 0.5 X_{2} + 0.6 X_{3} 
       - 0.4 U_{1} - 0.5 U_{2} - 0.6 U_{3}\bigr) \\
Y \mid \overline{A}_{3},\overline{\boldsymbol X}_{3},\overline{U}_{3} 
  &\sim \text{Normal} \ \!\Bigl(
       10 - 4 A_{3} - 3 A_{2} - 2 A_{1}
       -   X_{1} - 2 X_{2} - 3 X_{3} 
       +   U_{1} + 2 U_{2} + 3 U_{3},
       2\Bigr)
\end{align*}

\subsubsection{Two Time-varying unmeasured confounders, one continuous, one binary}

We simulated a data set that contains three visits, with $n$ subjects. For each subject $i$, at visit $j$, $j=1, 2, 3$, we simulated one binary time-varying unmeasured confounder $U_{ij}^{(1)}$ and one continuous time-varying unmeasured confounder $U_{ij}^{(2)}$, one binary time-varying measured confounder $X_{ij}$, and a binary time-varying treatment $A_{ij}$, and a continuous outcome $Y_{i}$. 

\begin{align*}
U^{(1)}_{1} &\sim \text{Bernoulli}(0.5) \\
U^{(2)}_{1} &\sim \text{Normal}(0.5,\,2) \\
X_{1} &\sim \text{Bernoulli}(0.5) \\
P(A_{1}=1 \mid U^{(1)}_{1},U^{(2)}_{1},X_{1}) 
  &= \operatorname{logit}^{-1}\!\bigl(0.5 X_{1} - 0.5 U^{(1)}_{1} - 0.2 U^{(2)}_{1}\bigr) \\
P(U^{(1)}_{2}=1 \mid U^{(1)}_{1},A_{1}) 
  &= \operatorname{logit}^{-1}\!\bigl(-0.5 U^{(1)}_{1} - 0.5 A_{1}\bigr) \\
U^{(2)}_{2} \mid U^{(2)}_{1},A_{1} 
  &\sim \text{Normal}\!\bigl(-0.5 U^{(2)}_{1} - 0.5 A_{1},\,2\bigr) \\
P(X_{2}=1 \mid X_{1},A_{1}) 
  &= \operatorname{logit}^{-1}\!\bigl(0.5 X_{1} + 0.5 A_{1}\bigr) \\
P(A_{2}=1 \mid U^{(1)}_{1},U^{(1)}_{2},U^{(2)}_{1},U^{(2)}_{2},X_{1},X_{2},A_{1}) 
  &= \operatorname{logit}^{-1}\!\bigl(1 - 0.3 A_{1}  + 0.4 X_{1} + 0.5 X_{2} \\
     &  - 0.2 U^{(1)}_{1} - 0.4 U^{(1)}_{2}
       - 0.1 U^{(2)}_{1} - 0.1 U^{(2)}_{2}\bigr) \\
P(U^{(1)}_{3}=1 \mid U^{(1)}_{2},A_{2}) 
  &= \operatorname{logit}^{-1}\!\bigl(-0.5 U^{(1)}_{2} - 0.5 A_{2}\bigr) \\
U^{(2)}_{3} \mid U^{(2)}_{2},A_{2} 
  &\sim \text{Normal}\!\bigl(-0.5 U^{(2)}_{2} - 0.5 A_{2},\,2\bigr) \\
P(X_{3}=1 \mid X_{2},A_{2}) 
  &= \operatorname{logit}^{-1}\!\bigl(0.5 X_{2} + 0.5 A_{2}\bigr) \\
P(A_{3}=1 \mid U^{(1)}_{1},U^{(1)}_{2},U^{(1)}_{3},
                   U^{(2)}_{1},U^{(2)}_{2},U^{(2)}_{3},
                   X_{1},X_{2},X_{3},A_{2}) 
  &= \operatorname{logit}^{-1}\!\bigl(1 - 0.3 A_{2} 
       + 0.3 X_{1} + 0.5 X_{2} + 0.6 X_{3} \\
      & - 0.2 U^{(1)}_{1} - 0.3 U^{(1)}_{2} - 0.4 U^{(1)}_{3}
       - 0.1 U^{(2)}_{1} - 0.1 U^{(2)}_{2} - 0.1 U^{(2)}_{3}\bigr) \\
Y \mid \overline{A}_{3},\overline{\boldsymbol X}_{3},\overline{U^{(1)}}_{3},\overline{U^{(2)}}_{3} 
  &\sim \text{Normal}\!\Bigl(
       10 - 4 A_{3} - 3 A_{2} - 2 A_{1}
       -   X_{1} - 2 X_{2} - 3 X_{3} \\
  &\quad + U^{(1)}_{1} + 2 U^{(1)}_{2} + 3 U^{(1)}_{3}
       + 0.1 U^{(2)}_{1} + 0.1 U^{(2)}_{2} + 0.1 U^{(2)}_{3},
       2\Bigr)
\end{align*}

\subsubsection*{No presence of unmeasured confounder}

Lastly, we simulated standard longitudinal causal data without unmeasured confounding with $n$ subjects and three visits. 
\begin{align*}
X_{1} &\sim \text{Bernoulli}(0.5) \\
P(A_{1}=1 \mid X_{1}) 
  &= \operatorname{logit}^{-1}\!\bigl(-0.1 + 0.2 X_{1}\bigr) \\
P(X_{2}=1 \mid X_{1},A_{1}) 
  &= \operatorname{logit}^{-1}\!\bigl(0.5 X_{1} + 0.5 A_{1}\bigr) \\
P(A_{2}=1 \mid X_{1},X_{2},A_{1}) 
  &= \operatorname{logit}^{-1}\!\bigl(-0.1 - 0.2 A_{1} + 0.1 X_{1} + 0.2 X_{2}\bigr) \\
P(X_{3}=1 \mid X_{2},A_{2}) 
  &= \operatorname{logit}^{-1}\!\bigl(0.5 X_{2} + 0.5 A_{2}\bigr) \\
P(A_{3}=1 \mid X_{1},X_{2},X_{3},A_{2}) 
  &= \operatorname{logit}^{-1}\!\bigl(-0.1 - 0.2 A_{2} 
       + 0.1 X_{1} + 0.2 X_{2} + 0.3 X_{3}\bigr) \\
Y \mid \overline{A}_{3},\overline{\boldsymbol X}_{3} 
  &\sim \text{Normal}\!\Bigl(
       10 - 4 A_{3} - 3 A_{2} - 2 A_{1}
       - 0.5 X_{1} - X_{2} - 1.5 X_{3},
       2\Bigr)
\end{align*}

\subsection*{2. Additional simulation results}

\begin{table}[H]
\centering
\begin{tabular}{|l|l|ccccc|}
\hline
Setting & Estimator & Mean & RB & SD & SE & CP \\ \hline
\multirow{6}{*}{\begin{tabular}[c]{@{}l@{}}n = 500, Time-Varying Bernoulli $U$\\ True ATE = -10.27\end{tabular}} 
& MSM $U$ included & -10.28 & -0.13 & 0.66 & 0.71 & 95.90 \\ 
& MSM $U$ excluded & -11.48 & -11.76 & 0.60 & 0.63 &  49.20\\ 
& Sensitivity Function (frequentist MSM) & -10.33 & -0.59 & 0.49 & 0.63 &   98.60\\
& Sensitivity Function (Bayesian MSM) & -10.34 &-0.64 & 0.49 & 0.57& 97.60\\ 
& BSA time-invariant $U$ & -10.58& -3.06& 0.76& 1.04& 97.10 \\ 
& BSA Time-varying $U$& -9.77 & 4.90 & 0.91 & 1.58 &  99.10 \\ \hline
\multirow{6}{*}{\begin{tabular}[c]{@{}l@{}}n = 500, Time-Varying Normal $U$\\ True ATE = -9.74\end{tabular}} 
& MSM $U$ included & -9.77 & -0.35 & 0.73 & 0.59 &  93.80 \\  
& MSM $U$ excluded & -10.46 & -7.35 & 0.38 & 0.44 &  63.0 \\
& Sensitivity Function (frequentist MSM)  &-9.69 &  0.52 & 0.32 & 0.44 & 99.40\\
& Sensitivity Function (Bayesian MSM)  &-9.69 &  0.52 & 0.31 & 0.38 & 98.10\\
& BSA time-invariant $U$ & -10.08 & -3.49 & 0.41 &0.63 & 98.10\\
& BSA Time-varying $U$ & -9.59 & 1.55 & 0.48 & 1.11 &  99.90 \\  \hline
\multirow{6}{*}{\begin{tabular}[c]{@{}l@{}}n = 500, two Time-Varying $U$s,\\ one Normal $U$, one Bernoulli $U$\\ True ATE = -10.40\end{tabular}}
& MSM $U$ included & -10.43 & -0.34& 0.74 & 0.76 & 94.90 \\  
& MSM $U$ excluded & -11.84& -13.88& 0.58& 0.61&  34.80 \\
& Sensitivity Function (frequentist MSM)  & -10.57&  -1.67 & 0.46 & 0.61  & 98.10\\
& Sensitivity Function (Bayesian MSM)  & -10.57 &  -1.70 & 0.46 & 0.55  & 97.10\\
& BSA time-invariant $U$ & -1.21 &88.17 &1.68 & 4.06 &40.0\\ 
& BSA Time-varying $U$ & -10.20 &1.87& 0.91& 1.58 & 99.3 \\  \hline
\multirow{6}{*}{\begin{tabular}[c]{@{}l@{}}n = 1000, Time-Varying Bernoulli $U$\\ True ATE = -10.27\end{tabular}}
& MSM $U$ included & -10.26 & 0.13 & 0.44 & 0.51 & 98.30 \\ 
& MSM $U$ excluded& -11.47 & -11.72 & 0.41 & 0.44 & 22.50 \\ 
&Sensitivity Function (frequentist MSM) & -10.33 & -0.54 &0.33 &0.45 &  99.60\\
& Sensitivity Function (Bayesian MSM) & -10.33 & -0.56 &0.33 &0.41 &  99.30\\
& BSA time-invariant $U$ & -10.48& -1.99& 0.82& 0.88&  93.10 \\ 
& BSA Time-varying $U$& -9.67 & 5.88 & 0.99 & 1.43 & 96.80 \\  \hline
\multirow{6}{*}{\begin{tabular}[c]{@{}l@{}}n = 500, Time-invariant Bernoulli $U$\\ True ATE = -9.62\end{tabular}} 
& MSM $U$ included& -9.63 & -0.19 & 0.45 & 0.45 & 94.39 \\ 
& MSM $U$ excluded& -10.63 & -10.56 & 0.43 & 0.45 & 38.24 \\ 
& Sensitivity Function (frequentist MSM) & -9.96 &  -3.63 & 0.36 & 0.45 & 91.90\\
& Sensitivity Function (Bayesian MSM) & -9.96 &  -3.65 & 0.36 & 0.44 & 89.40\\
& BSA Time Constant $U$ & -9.99 & -3.88 & 0.63 & 0.82 &  97.20 \\ 
& BSA Time-varying $U$ & -9.27 & 3.54 & 0.76 & 1.37 &  99.40 \\ \hline
\multirow{5}{*}{\begin{tabular}[c]{@{}l@{}}n = 500, $U$ doesn't exist\\ True ATE = -9.30\end{tabular}} & MSM & -9.31 & -0.08 & 0.36 & 0.35 & 93.90\\ 
& Sensitivity Function (frequentist MSM)  &-9.14  &1.84 &0.30 &0.37& 96.30\\
& Sensitivity Function (Bayesian MSM)  &-9.14  &1.84 &0.30 &0.34 & 95.10\\
& BSA time-invariant $U$ & -9.04& 2.85 & 0.41& 0.64 & 98.80\\ 
& BSA Time-varying $U$ & -8.74 & 6.08 & 0.50 & 0.97  & 99.20 \\ \hline
\end{tabular}
\caption{\ \ Simulation results for the estimated causal parameter $E[Y(1,1,1)] - E[Y(0,0,0)]$ over 1000 replications among different settings. The columns are simulation settings including sample size, varying distribution of unmeasured confounder, different forms of unmeasured confounder, posterior predictive mean, relative bias (RB), standard deviation (SD), standard error (SE), and 95\%coverage probability (CP). MSM: marginal structural models; BSA: Bayesian Sensitivity Analysis approach.}
\label{tab:long}
\end{table}

\subsection*{3. Additional analysis results}

The majority of the patients were male (58.1\%) with the average age at baseline of 12.3 years (4.1 SD). In addition, 57\% of patients were diagnosed with IBD before PSC onset. We observed some differences between OVT exposure groups in sex, concurrent IBD diagnosis, baseline GGT and APRI, and baseline exposure to other medications including steroids and aminosalicylates, indicating an imbalanced baseline profile.

\begin{table}[h]
\centering
\begin{tabular}{rllll}
  \hline
    \hline
 & Overall & OVT = 00 & OVT = 01 & OVT =  11 \\ 
 &   $n = 401$ & $n = 343$ &  $n =  19$ &  $n =  39$ \\ 
 \hline

 \textbf{Demographic}\\
  Male (N(\%)) &   233 (58.1)  &   196 (57.1)  &    10 (52.6)  &    27 (69.2)  \\ 
  Age (mean (SD)) & 12.27 (4.08) & 12.15 (4.24) & 12.58 (3.34) & 13.14 (2.68) \\ 
  
\hline
\textbf{Baseline clinical measurements}\\
  Large duct phenotype (N(\%)) &   358 (89.3)  &   307 (89.5)  &    17 (89.5)  &    34 (87.2)  \\ 
  Concurrent IBD (N(\%)) &   229 (57.1)  &   191 (55.7)  &    14 (73.7)  &    24 (61.5)  \\ 
  Overlapping AIH (N(\%)) &   119 (29.7)  &    99 (28.9)  &     6 (31.6)  &    14 (35.9)  \\ 
 Metavir Stage III-IV (N(\%)) &    84 (20.9)  &    74 (21.6)  &     1 ( 5.3)  &     9 (23.1)  \\ 
Hepatobiliary Event (N(\%))&    38 ( 9.5)  &    32 ( 9.3)  &     0 ( 0.0)  &     6 (15.4)  \\
\hline
\textbf{Baseline Biochemistry}\\
GGT (median [IQR]) & 47 (19-153)& 52 (19-153.5) & 77 (27-193) & 33 (16-103) \\ 
  $ln(GGT)$ (mean (SD)) &   4.03 (1.25) &  4.04 (1.25) &  4.16 (1.20) &  3.82 (1.28) \\ 

\hline
\textbf{Other Medications}\\
  UDCA (N(\%)) &   145 (36.2)  &   131 (38.2)  &     6 (31.6)  &     8 (20.5)  \\ 
  Steroids (N(\%)) &    76 (19.0)  &    60 (17.5)  &     6 (31.6)  &    10 (25.6)  \\ 
 Immunomodulators (N(\%)) &    88 (21.9)  &    73 (21.3)  &     5 (26.3)  &    10 (25.6)  \\ 
   \hline  \hline
\end{tabular}
\caption{\ \ Baseline characteristics by OVT exposure.OVT: oral vancomycin therapy; OVT=00:OVT free within 2 years; OVT=01: OVT free between 0 to 1 year and exposed between 1 to 2 years; OVT=11:OVT exposed between 0 to 1 year and remained on OVT between 1  to 2 years; GGT: gamma-glutamyl transferase;  IBD: inflammatory bowel disease; AIH: autoimmune hepatitis; UDCA: ursodeoxycholic acid; SD: standard deviation.\label{da: Table_demo}}
\end{table}

\end{document}